# Two thousand years of the oracle problem. Insights from Ancient Delphi on the future of blockchain oracles.

Giulio Caldarelli[a,*], Massimiliano Ornaghi[b]

[a,b]University of Turin

[a]Department of Management, Corso Unione Sovietica 218bis, 10134, Turin To, Italy
[b]Department of Humanities, Via Sant'Ottavio, 20, 10124, Turin To, Italy

*Correspondence address: giulio.caldarelli@unito.it

**Abstract**

The oracle problem refers to the inability of an agent to know if the information coming from an oracle is authentic and unbiased. In ancient times, philosophers and historians debated on how to evaluate, increase, and secure the reliability of oracle predictions, particularly those from Delphi, which pertained to matters of state. Today, we refer to data carriers for automatic machines as oracles, but establishing a secure channel between these oracles and the real world still represents a challenge. Despite numerous efforts, this problem remains mostly unsolved, and the recent advent of blockchain oracles has added a layer of complexity because of the decentralization of blockchains. This paper conceptually connects Delphic and modern blockchain oracles, developing a comparative framework. Leveraging blockchain oracle taxonomy, lexical analysis is also performed on 167 Delphic queries to shed light on the relationship between oracle answer quality and question type. The presented framework aims first at revealing commonalities between classical and computational oracles and then at enriching the oracle analysis within each field. This study contributes to the computer science literature by proposing strategies to improve the reliability of blockchain oracles based on insights from Delphi and to classical literature by introducing a framework that can also be applied to interpret and classify other ancient oracular mechanisms.

1. Introduction

"*If you go to war against Persia, you will destroy a great empire*" (Herodotus, 1.53). This famous prophecy from Delphi, which led Croesus, King of Lydia, to lose his kingdom, was both revered and criticized for its ambiguity and its potential to be misinterpreted (Bowden, 2005). Ancient historians and philosophers, many of whom accepted that oracles were messengers from the gods, nonetheless debated the reliability of oracles, because they recognized that oracles were susceptible to human influences. Oracle-seekers, particularly wealthy or powerful ones, could use bribery or influence to obtain an oracle to achieve their desired outcome. They could unintentionally misremember an oracle to suit their needs or misinterpret an oracle, particularly if it was vague or ambiguous, as Croesus did (Fairbanks, 1906). A false or manipulated prediction, as well as a misinterpretation of a genuine one, could lead to an unwanted outcome.

The ancient appreciation of oracles, especially Delphic ones, which were famous for their ambiguity as sources of both guidance and confusion, has a parallel with the "oracle" in computer science, a term introduced by Alan Turing in (1939). The oracle machine was a Turing machine with access to a black box able to answer specific questions instantly. The idea was to push the boundaries of what a Turing machine could solve by letting it rely on an external source of information. Still, the whole model depended on assuming that the oracle's answers were correct. If they were not, the output would also be unreliable (Arkoudas, 2008). The rise of decentralized ledger technologies (DLTs), such as blockchains, which rely on oracles to fetch external data, has brought these concerns back into focus.

In its simplest form, an oracle is an intermediary between the smart contract and real-world data, just like the priestess of Delphi was an intermediary between humans and the god Apollo. Since Turing machines are closed systems, they have no way of communicating with the real world, so the oracle machine allows an automatic machine to fetch real-world data. In ancient times, humans sought knowledge from the gods beyond their understanding or about future events. The oracle was the intermediary that allowed information from the world of gods to be transmitted to humans who had no access to it. Figure 1 exemplifies this parallelism.

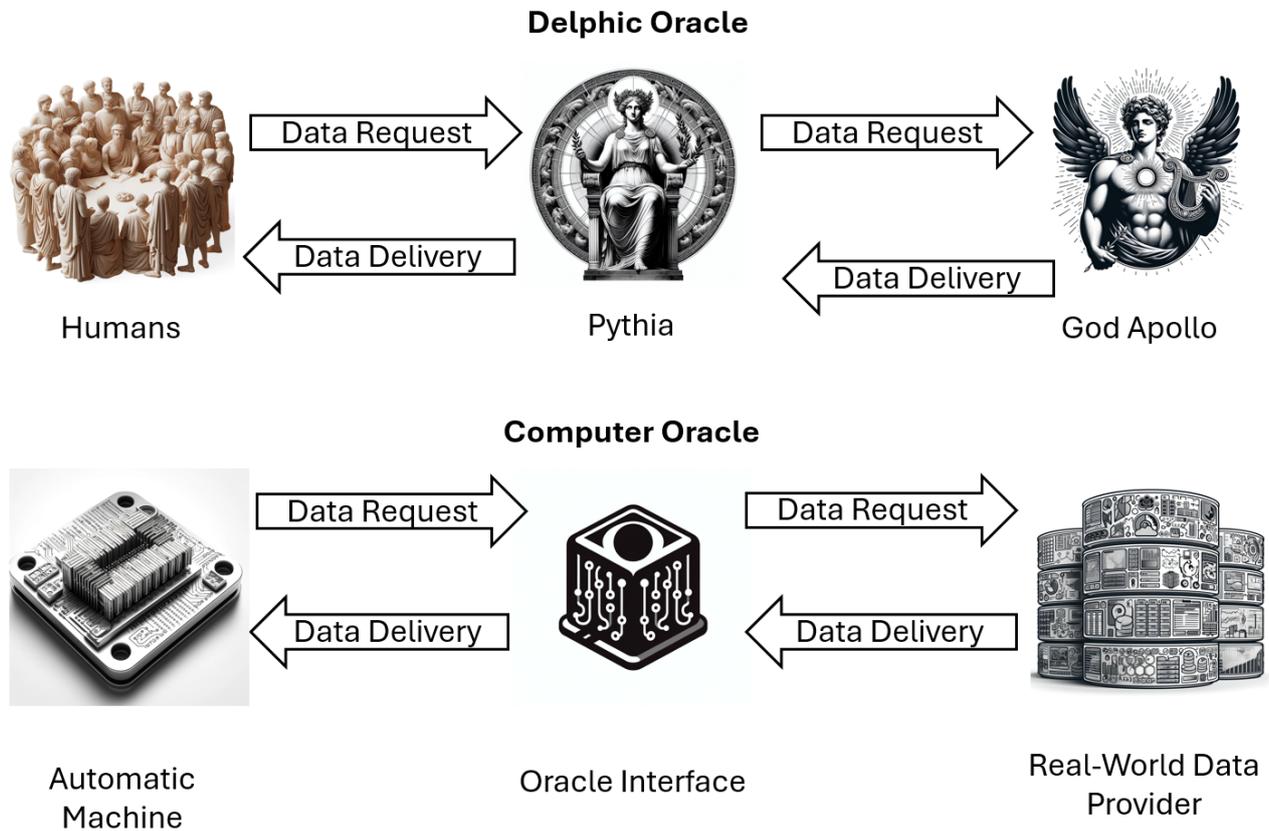

**Figure 1.** Parallelism between Delphic and computer oracles.

Many authors and practitioners then proposed new oracle solutions to improve the reliability of data transfer and reduce the chance of manipulation. However, despite leveraging the most advanced technologies, oracles still suffer from attacks and manipulation.

We believe that the oracle problem is not being solved by advancements in technology because it is not strictly tied to a lack of technology. We believe the oracle problem is a more philosophical concept whose solution, as already proposed in (Caldarelli, 2020b) requires the cooperation of experts from different disciplines. Examples of oracles, such as Truthcoin based on Game Theory, that inspired many other robust oracle solutions, support this idea (Sztorc, 2015).

The scope of this study is to create a layer of abstraction for blockchain oracles and the oracle problem that can be examined across disciplines. Historians, philosophers, economists, and computer scientists alike may contribute to a fuller understanding of the issue, showing that oracle-related questions benefit from a broad scholarly perspective. We acknowledge prior studies that already contributed to generalizing aspects of the oracle problem by underlying the patterns and common challenges (Caldarelli, 2020b; Mühlberger *et al.*, 2020; Eskandari *et al.*, 2021; Hassan *et al.*, 2023). However, to the best of our knowledge, a thorough comparison has not been made between the blockchain oracle problem and the ancient divination oracle designs. Poblet et al. (2020) already made some parallelism between Athen's and blockchain oracles, but as a dispute resolution mechanism for smart contracts. Instead, the idea of the paper we present, rather than merely comparing these two types of oracles, serves to further enlighten that oracle design and challenges have not been substantially modified throughout the eons of time.

To achieve our study goal, we first provide an overview of prior studies on blockchain oracles and the oracle problem that already offered a generalization of the issue, creating an optimal framework for comparison. Then, we compare these designs and challenges with information concerning ancient Delphic oracle designs drawn from the most renowned and actual resources. We then discuss common patterns and designs emerging from the study, proposing multidisciplinary avenues of research. Although there are countless ancient oracles, we decided to focus on the Delphic one for the following reasons. First, most of the available study material concerns the Delphic oracle; therefore, it allows for a better comparison between the two. Second, being the most renowned and famous, it is arguably better than others, and its characteristics are worthy of investigation and comparison with modern oracles in order to suggest possible improvements.

The paper proceeds as follows. Section 2 introduces the literature background, while Section 3 explains the methodology chosen for the research. Section 4 outlines the outcome of the comparison, while Section 5 discusses the findings. Section 6 concludes the paper, providing further avenues for research.

## 2. Literature Background

### 2.1 Ancient oracle's designs and limitations.

The etymology of the word "oracle" derives from the Latin "*oraculum*," which in turn comes from "orare," meaning "to speak." Oraculum refers to a "divine announcement" and denotes the act of a third party reporting the word or prophecy issued by a god or divine entity. In light of this definition, many historians accurately regard as oracles those divination mechanisms that involve an intermediary between the petitioner and the source of divine knowledge. For instance, the practice of *Augury*, which involved interpreting the will of the gods through the flight patterns of birds, among other omens, can hardly be considered an oracle mechanism. In this case, in fact, the divine entity is not directly questioned, but its will is indirectly interpreted (Beard, North and Price, 1998). Similarly, *Astrology* is not classed as an oracular mechanism, as it relies on the presumed correlation between the deterministic motion of planets and human destiny.

The most classic example of an oracular mechanism is a divination temple (e.g., Delphi or Dodona), where a priest or priestess delivers responses on behalf of a divine entity. What sets an oracle apart from an ordinary omen is precisely this act of intermediation.

Perhaps the most influential of all ancient oracles, was the Delphic Oracle, situated on the slopes of Mount Parnassus, consulted by all strata of Greek society and beyond, from peasants to kings. The Pythia, a priestess of Apollo, delivered cryptic predictions while in a trance-like state, according to some historians (Foster and Lehoux, 2007) induced by natural gases emanating from the earth. These pronouncements could sway public policy and personal decisions significantly (Fontenrose, 1978).

The rationale behind these oracular practices stemmed from a belief in the active involvement of gods in human affairs, with oracles serving as a tangible medium through which divine guidance could be solicited. The ambiguous nature of many oracular pronouncements, however, opened the door to manipulation by those seeking to use divine authority to validate their personal desires or political ambitions. Historical records suggest instances where interpretations were strategically molded to support specific outcomes, demonstrating the interplay between divine consultation and human agency (Maurizio, 1995).

Consulting the Oracle at Delphi was quite an articulate process, depending on the petitioner's identity. As the temple was located in Delphi, it was available only to its citizens or, in rare cases, to those "sponsored" by one of them. On a specific day of the month, when the temple was open to petitioners, they had to first purify themselves, then offer a goat in sacrifice, and then they could enter the temple and ask their questions. The priestess of the temple, the Pythia, after washing herself in the Castalian Spring and entering the temple, would ask the question of the god Apollo and communicate the god's response to the petitioner. The process of consulting the oracle for cities was instead a bit more complex. First, the city could not query any oracle but only the available ones unless a special right was granted to consult others. Second, oracles were subject to "seasonality" therefore, not all oracles were always available. Third, as the medium/priest could not leave the sacred temple, he was unable to speak directly to the council, the assembly or any other entity that requested the predictions. In order to query the oracle, it was necessary then to send some "trusted" emissary to the temple. Furthermore, as the medium was not allowed to write (most probably because it was not instructed enough), the prediction had to be transcribed on paper by the trusted emissary. The message had then to be brought safely to the assembly, and the meaning of the message had to be straightforward for them to understand (Bowden, 2005). Intuitively, a mechanism like this was not infallible. Being the voice of the god, oracular responses were rarely questioned. The first doubts about the legitimacy of divination emerge already in the classical period, with authors such as Euripides and Xenophon showing more pragmatic or critical attitudes toward divine signs (Xenophon, Memorabilia 1.1.9; Euripides, Ion 418-455). Later, in the Hellenistic period, Academic and Epicurean philosophers developed more systematic critiques of oracles and divination (Cicero, De divinatione 2.12). However, before that specific era, Herodotus and Thucydides already criticized not the oracle itself as the god's emissary but the way oracular responses were obtained (Fontenrose, 1978; Plutarch, Moralia V). The above-mentioned mechanism of consulting the oracle, in fact, could fail in many ways. Taking for granted that the oracle always spoke the truth, there was still the possibility that the trusted intermediary failed to deliver the correct message. The assembly could receive the wrong message, either because the intermediary misunderstood the message itself or because it was corrupted. As the oracle's decision was not debatable, the manipulation of its message meant a manipulation of the assembly itself. When assembly decisions concerned matters of high importance, also high was the chance for them to be manipulated. Herodotus and Thucydides both describe a story in which Athenians corrupted the Pythia to influence the Spartans (Bowden, 2005)

Thus, while ancient oracles were revered as sacred channels to the divine, their influence was not immune to the machinations of human intent, reflecting the complex relationship between faith, power, and manipulation in the ancient world.

**2.2 Blockchain oracles**

In the original concept of Nakamoto's contracts, oracles were not needed, as any signer had to be aware of and agree with the contract data. Once the required number of signers agreed to a condition or to an update, the code written in the contract was executed at its expiration. Therefore, no specific entity was in charge or required to provide data to the contract (Nakamoto, 2010). This contract design, however, made their execution time demanding and difficult to apply in a real-world scenario. Therefore, an early Bitcoin developer, Mike Hearn, proposed to feed a smart contract with data coming from a trusted third party leveraged as an "oracle" (Hearn, 2011). For example, in the case of a testament contract, an external party had the task of communicating the event of death so that the inheritance could be released to the beneficiary. After the simplest blockchain oracle mechanisms of 2011, many different designs and ideas were generated (Caldarelli, 2023). Several works of literature proposed a systemization of knowledge for oracle design and schemes that, although not exhaustive, are useful to have a broad understanding of their general architecture (Mühlberger *et al.*, 2020; Eskandari *et al.*, 2021).

**2.3 Oracle architecture.**

Apart from their design that may vary, Oracle mechanisms are generally composed of three parts. The data source, the communication channel and the contract.

1) The data source is the actor, database or IoT, that provides a certain piece of information. It's the sole responsible of the genuineness and reliability of the data. Data provider honesty is a necessary condition but not sufficient for the oracle to be reliable.
2) The communication channel is the means through which the data is brought to the blockchain. In certain cases, such as for human oracles, the data source is also the communication channel that writes the data into the smart contract. The responsibility of a communication channel is not only to provide the data from the source to the contract but also to ensure that the data is not manipulated during transmission or altered prior to its digestion.
3) The contract is the software that technically digests the data and executes the operation for which the data feed was needed. Intuitively, it's vital that the contract is well written and free of bugs, but it's also the contractor's responsibility to make sure that the data requested is appropriate for the contract purpose. Making sure that real-time data or time-weighted-averaged data is appropriate for the purpose of the application is, for example, the responsibility of the contract.

For an oracle to work properly, all these parts need to operate perfectly and be combined wisely. Intuitively, the data source should not only be reliable but appropriate for the intended purpose. Similarly, the communication channel should be secure to allow data to arrive at the contract securely, preventing unwanted manipulation. It may also need to guarantee some degree of privacy depending on the purpose of the smart contract. Finally, the contract, as specified, should not only guarantee the absence of bugs but should also be properly written. This means that it has to point to the data source adapted for the purpose, and it has to properly digest its data, performing the required computations if necessary. Any type of failure or deficiency in this complex architecture may dramatically alter the reliability of the oracle.

**2.4 Oracle types.**

To date, oracle solutions are countless, but we can divide them in two main categories, brilliantly described in Heiss et al. (2019).

1) **Transport Layer Security-based:** Are oracles whose main purpose is to ensure that the data digested by the contract is authentic and genuine. Often labeled as "centralized" solutions, those oracles tend to sacrifice decentralization in order to maximize efficiency and security. Their architecture allows leveraging enclaves such as Trusted Execution Environments (TEEs) that provide an attestation that the data is authenticated, preventing any unwanted access to it during its transport. These types of oracles are mainly appropriate when the data is not in the public domain, or reliable data sources are scarce, or else when there is a need to reduce costs or guarantee a higher level of data security.

2) **Voting-based:** often labeled as decentralized, these oracles leverage multiple data sources. Oracles of this kind, after gathering data from all the available sources, select the appropriate piece of data to finally send to the smart contract. These oracles are used when data is publicly accessible and there is a chance that a specific data source may be manipulated or unavailable. The criterion to digest the data typically varies according to the specific protocol and its real-world application. When obtaining diverging pieces of data from multiple sources, it can be averaged if numeric; otherwise, a voting mechanism can decide the piece of data considered as most appropriate for the contract purpose. Often, in specific circumstances, these oracles are used when the requested data is not available anywhere, and a triangulation of multiple data sources is thought to generate the requested piece of data.

We wish to stress that these distinctions based on centralization, as well as others, are purely for didactical and speculative purposes. Decentralized oracle solutions are, for example, usually centralized in nature, with the contract managed solely by the company that manages the protocol. On the other hand, centralized oracle solutions can be decentralized if managed by a decentralized autonomous organization (DAO) that democratically elects appropriate data sources or communication channels. However, for research purposes and in particular for the theoretical aspect of this specific research, a general distinction had to be made to better understand the parallelism with classical oracle schemes.

**2.5 Oracle challenges and risks.**

The so-called oracle problem implies the dependencies of a trustless system, such as the blockchain, on a non-trustless element, such as an oracle. This condition entails several different challenges and risks. Prior research (Heiss, Eberhardt and Tai, 2019; Al-Breiki *et al.*, 2020; Pasdar, Dong and Lee, 2021), brilliantly provides a standardized overview of these. Given the focus of the present study, very technical, related characteristics and challenges are not presented.

**Attributability**: The data provided to the contract must be attributable to a specific data source, which should be known.

**Accountability**: The source of the information given to the oracle should be putting something at stake for providing false or incorrect information.

**Authenticity**: It should be possible to verify the authenticity of the data received from the data source.

**Integrity**: It should be possible to verify that the data received has not been altered from the information originally submitted by the data source.

**Availability**: Data can be retrieved whenever it is necessary.

**Latency**: Regards the time it takes from the query to be received by the oracle to the response to be given to the requester.

**Accessibility**: No barriers to access are imposed on the requester. Costs are, for example, a known barrier that may significantly reduce the accessibility of an oracle. Other oracles may only be accessible to some individuals and, therefore, are not open to everyone.

Depending on the degree of independence that the oracle has, the following risks can be encountered:

**Centralization**: The entire oracle mechanism is under the sole control of a single entity, which can arbitrarily select the data source and monitor the communication channel. This creates a single point of failure.

**Collusion**: In a strongly centralized environment, the authority managing the oracle may collude with an external party to manipulate the oracle outcome for selfish interests.

**Sybil attacks**: In a decentralized oracle network where nodes are anonymous, a single party may impersonate multiple nodes to manipulate the outcome of a democratic choice.

**Lazy Equilibrium**: Equilibrium may be reached with nodes that provide standardized answers to different questions, without performing any data validation.

**Freeloading**: Nodes simply copy and broadcast what other oracles suggested as reliable data, de facto, without having the effort of verifying if the data provided by other sources is reliable.

**2.6 Queries**

An oracle is fundamentally a device that answers questions, although not every fact may be considered in the form of a question. Despite the crucial importance of queries, only a couple of papers discuss them. Pasdar et al. (2021), distinguish queries in binary, scalar, and categorical, unfortunately, without further delving into their distinction. Bartholic et al. (2022), on the other hand, provide a thorough analysis of oracle queries and challenges emerging from each type. A detailed taxonomic analysis is provided, which is summarized as follows.

**Events**: queries logical statements, which can be true, false, or unknown by the oracle.

**Non-Events**: Queries whose outcome is purely random and therefore can not be answered by any oracle. In this category are included generic questions that do not query over a specific event.

Events can be further distinguished in:

**Recondite Events**: Queries that can be answered only by a specific data source and are unknown to the public.

**Sanctioned Events**: Queries that can be public or not, but for which a subset of parties is able to answer authoritatively.

**Discernible Events**: Concerns events that can be broadly observed and for which a large number of parties can answer related queries.

**Computational Events**: Involve queries that can be answered by performing a logical computation without specific limitations on any party computing the answer.

**Ambiguous Events**: Concerns queries that are answered differently by honest and informed parties. The divergence may be due to the available information for each answering party, or to different views of reality or of the event context in query framing.

Queries involving non-event or ambiguous events are intuitively undesirable; however, literature further specifies that although queries are of heterogeneous types, the performance of oracles excels when the range of possible answers is limited (Bartholic *et al.*, 2022). To make an example, a query about the color of the sky may be framed in an open ended way (e.g., what color is the sky?), including all existing colors as possible answers or may be framed as binary question (e.g., Is the sky blue?), indeed excluding all answers that differ from yes, no, and unknown. This narrow area of research is particularly important as it clarifies that a crucial variable that affects the reliability of an oracle performance is the structure and type of query.

The specific design that is sought to balance optimal oracle characteristics while reducing potential barriers is called the "trust model" (Hassan *et al.*, 2023). To date, various blockchain oracle solutions have multiple trust models, but despite their market capitalization dominance, none has undisputed dominance over the others. In ancient times, the Oracle at Delphi gained significant dominance over other oracle prediction mechanisms and sources of information. This may not necessarily imply overall superiority over other oracles, but it definitely supports its unique design. A blockchain oracles that mirrors Delphic characteristics may also present unique and useful features.

3. **Methodology**

The study's methodology is outlined as follows. The first part provided in Chapter 2 consisted of extracting blockchain oracles' characteristics, challenges, and risks from the current literature. As noted in previous studies (Caldarelli, 2020a, 2022) research on the subject is very narrow, and studies involving oracle classifications are a few and well-known. Furthermore, aiming to keep the analysis at a theoretical level, classification based on specific hardware was excluded. Once the list of characteristics is drawn, an analysis is made in the second part of this study to understand how the oracle of Delphi positions itself toward these characteristics. Characteristics, as well as risks and challenges, are then extensively described one by one. As for the source of Delphic oracular mechanisms, two main repositories will be utilized, the Fontenrose (1978) and Parke and Wormell (1956). Although the authors recognize more recent studies on oracular mechanisms, these books are still considered the most complete source of information on the Delphic oracle. However, a discussion of an alternative piece of literature is made when analyzing specific characteristics. For each of the described characteristics, a summary table in the findings section provides parallels with blockchain oracles to better grasp the rationale of the study.

Finally, concerning queries, the analysis implements a quantitative approach. In this last part, the idea is to understand how the Delphic oracle answers based on the question types, drawing parallels with query types classification used in computer science. Being more recent and including more Delphic queries, Fontenrose (1978) is leveraged as the main data source for this research phase. We selected Historical and Quasi-historical queries, as other types are known to be fictional, so they will not genuinely represent the Delphic answering mechanism. Incomplete queries or queries missing either question or answer were discarded. The remaining 167 query pairs were classified according to Bartholic (2022) classification of blockchain oracle queries. We understand that Delphic queries classification we performed may present a certain level of subjectivity, but we provide the complete table with classification criteria so that the study is openly reproducible. Once queries are classified, answers are inspected using various techniques, leveraging Python libraries such as TextBlob and SciPy. Word count is used to have a general idea of the verbosity of the responses, while Shannon Entropy should give an idea of their complexity. Shannon Entropy (1 to 5) measures the occurrences of unusual words, measuring the complexity of a sentence. Given the known complexity of Delphic responses, we expect high values on averages but with peaks for more blurry categories. Modal density shows instead the authoritative tone of the answer, while hedge density shows the degree of uncertainty. Polarity displays the emotional tone of an answer, ranging from -1 (negative) to 1 (positive), with a neutral value close to 0. Subjectivity finally displays the extent to which the text expresses personal judgment, ranging from 0 (fully objective) to 1 (fully subjective). The rationale for this classification is to understand if and how the answer given by the oracle is influenced by the given query. Drawing parallels with blockchain oracles, the findings of this section will allow us to understand under what type of query a blockchain oracle will provide the most reliable responses. The next section introduces the findings starting from the Delphic consultation procedures.

### 4. Delphic oracle procedures, risks, and consultation queries.

The reconstruction of the history and functioning of the Oracle of Apollo at Delphi, from its origins (or at least from its institutionalization) until the Hellenistic period, is hampered by serious gaps in our textual documentation, both literary and epigraphic, with regard to the detailed description of the sanctuary's organizational procedures. This gap is both synchronic (the data, in general, are few and sporadic for each phase of the sanctuary's life) and diachronic (the data, for specific events or codified actions, are limited and scattered across different periods). Therefore, any reconstruction inevitably relies on information obtained from relatively late sources, which sometimes contain more extensive descriptions of Delphic procedures: a prime example is Plutarch (1st–2nd century CE), who devoted a group of pamphlets to Delphic themes.

This information is inevitably encrusted with centuries of interpretations (and perhaps misunderstandings), and therefore ideological superstructures. However, in order to produce even provisional working hypotheses, scholars have compared it with details preserved in earlier literary sources that contain references or allusions to the Delphic Oracle, the contexts of some questions, the questioners, and the temple officials involved. These sources include the Homeric Hymn to Apollo (henceforth, *HHAp.*: abbreviations of ancient sources follow *LSJ*, with some simplifications), Herodotus' *Histories*, and tragedies such as Euripides' *Ion* and Aeschylus' *Eumenides*. On the other hand, the possibility of comparison with material sources (especially archaeological ones) is limited, since these provide little information about ritual procedures.

Nevertheless, when the accounts of the earliest sources coincide with later and more detailed data (e.g. Plutarch), scholars tend to make some generalizations about how the Oracle operated. The obvious limits of these deductive procedures, therefore, require a very cautious approach when formulating hypotheses about the history and functioning of the Oracle, and they highlight the hypothetical nature of any statement that can be made. These precautions should also be kept in mind when approaching the considerations developed in the following pages (especially those based on isolated literary evidence).

As for the sources and, more broadly, the operation of the Delphic Oracle interpreted in relation to its history, the most balanced treatment to date is probably the monograph by Parke and Wormell (1956, following Parke's earlier history of the sanctuary [1939]), together with Amandry's foundational research (Amandry 1950). These studies remain the primary reference for further discussion of the issues raised in this text and can usefully be integrated, and in some respects, updated, with distinctive perspectives, by the more recent works of Fontenrose (1978) and Bowden (2005, esp. pp. 12-39, and 65-87), as well as by the monographs and papers of Delcourt (1955, now with Giangiulio's 2024 historiographical assessment), Vernant (1974, for a productive problematisation of the evidence), Lloyd-Jones (1976), Compton (1994), Catenacci (2001), and Mari (2017, esp. pp. 662 ff.).

### 4.1 Attributability

According to tradition, in historical times, the sanctuary of Delphi was consecrated to Apollo, who had his temple built there and founded his Oracle, as recounted in the *Homeric Hymn to Apollo* (*HHAp.* 294-299; on the three Delphic temples of historical times, that housed the oracle, see now Rougemont 2013). This episode, preserved in the second part of the hymn, most likely reflects a local tradition from Phocis (the region of the sanctuary), or at least from the mainland, and was therefore probably shared by the inhabitants of the sanctuary's area. The elements of the hymn attributable to its patron, probably from Samos (and thus non-local), do not appear to concern the mythical component that is the subject of our discussion (cf. Burkert, 1979; Aloni, 1989).

There were several oracular centres dedicated to Apollo, for example, the regional, if not local, Oracle of Ptoion, and the major sanctuaries of Klaros and Didyma (Miletus) in Ionia (cf. e.g. Fontenrose, 1988; Greaves, 2012). There is also a long-standing debate about the existence of an Oracle in the Apollonian sanctuary of Delos, in the Cyclades (cf. *HHAp.* 79-81, with Càssola 1975, pp. 87-88; Richardson 2010, pp. 93-94). The spread of these oracular centres dedicated to Apollo, among which Delphi eventually came to stand out, was due to the fact that prophecy was one of the god's principal "*timai*" (τιμαί), that is, one of the spheres or prerogatives considered proper to him: «To me shall be the lyre and the curved bow», says the newborn god in the Homeric hymn, «and I shall divine for men the unfailing thought of Zeus» (*HHAp.* 131–132; cf. also Pind. *Pyth.* V 63–69; Plat. *Crat.* 405a; Callim. *in Apoll.* 42–46). Apollo could read Zeus' mind and "communicate" the knowledge he derived from it, adopting different procedures at each of his oracles.

The prophetic responses given at Delphi were considered particularly authoritative: «Golden Pytho of the famous prophecies», says Pindar at the beginning of one of his songs (cf. fr. 52f Maehler, vv. 1–2). However, we know little about the early prophetic procedures of Apollo at Delphi: we have only vague references to an ancient prophetic laurel tree (cf. *HHAp.* 395–396, with Parke and Wormell 1956, Vol. I, pp. 3 ff. [esp. 26], and also Amandry 1950, pp. 126–134), or to practices of ornithomancy, which may have been merely preliminary to the actual consultation of the god (cf. *e.g. HHHerm.* 543–549: on the different divination procedures, in general, see also Amandry 1950, pp. 19 ff.; Catenacci 2001, pp. 136–167 [and note 17]).

In historical times, however, the answers provided in the sanctuary came from the inspiration that Apollo instilled in a woman called "Pythia" (cf. Maurizio 1995; Catenacci 2001, pp. 144 ff.; Pisano 2014; and already Dodds 1951, pp. 109 ff.). Thanks to the Pythia, the god's message was verbalised, usually in response to a question posed to the god by an interlocutor, or by his representative, or by representatives of a community. The Pythia acted as the mouthpiece or intermediary of the god and, almost always in myth, communicated his response directly to the person who asked the question. Historical sources, on the other hand, often depict the Pythia supported by officials of the sanctuary who acted as intermediaries, transferring the response to the questioner and, in some cases, probably transcribing it (cf. Amandry 1950, pp. 149 ff.; also §4.9).

We know, more specifically, that there was a priest (or perhaps more than one) in the sanctuary, but sources also mention a group of officials called "*prophetai*" (προφῆται), and it cannot be ruled out that the priest himself was one of these *prophetai* (and that this was a generic term used for the officiants of the temple and oracle). Later evidence, moreover, seems to indicate the existence of another group of officials called "*hosioi*" (ὅσιοι), whose functions are still unclear (see Amandry 1950, pp. 118-125; Compton 1994, p. 222; different opinion in Bowden 2005, p. 16).

The oracles were in prose or, perhaps more frequently, in verse, as seems to be the case especially in sources of the Archaic and Classical periods (cf. also §4.11). They were often composed in dactylic hexameters, *i.e.* the metre of epic poetry (cf. Parke, 1945). It is not clear, however, whether this form of expression depended on the Pythia herself or on the priests or officiants who conveyed the message to the questioner (cf. Parke and Wormell 1956, Vol. I, pp. 33–34; Bowden 2005, pp. 33–38, *e.g.* p. 38: «The verses produced at Delphi were therefore not intended to falsify the record of consultations, but to give them a grandeur that the actual spoken words might not have had»). Even in antiquity it was understood that this was the human component of the oracular message: according to Plutarch, for example, the message came from the god, while the diction (*lexis*, λέξις) and metre (*metron*, μέτρον) were from the Pythia (cf. Plut. *de Pythiae orac.* 397c). When the Pythia's response was expressed in the first person ("I"), people believed that it was the god himself speaking: "I" was Apollo, not the Pythia (cf. Parke and Wormell 1956, Vol. I, p. 40; cf. also §4.7, on the possession of the Pythia). In any case, confidence in the truthfulness and effectiveness of the god's word was not questioned, at least in the Archaic and Classical ages.

It is possible, however, that over time the number of bearers of the god's voice (*i.e.* the Pythia) varied. If we stick to Plutarch's account, the sanctuary of Delphi at the height of its splendour housed two Pythias, to which a third was added as *éphedros* (ἔφεδρος, "who sits next to", *i.e.* "who waits her turn", therefore "a spare"). That said, we do not know whether Plutarch, in mentioning the "splendour" of the Oracle, referred to the Archaic period, late archaism, or the Classical age; in any case, in Plutarch's own time (the Imperial Age), there was once again only one Pythia, as was probably originally the case (cf. Plut. *de defectu orac.* 414b). On the other hand, it cannot be ruled out that even in the early stages of the sanctuary's history, when the institution of the Pythia had been consolidated, there had already been established «a whole guild of consecrated women of mature years, who served in the temple, and who would provide a natural recruiting ground for the post of Pythia» (Parke and Wormell 1956, Vol. I, p. 36), purely for practical reasons, to safeguard the sanctuary's procedural mechanisms.

### 4.2 Accountability

As a clear result of what has just been said, a Delphic prediction could not be wrong, at least at its origin: Apollo could not make a mistake (as the divine holder of prophecy, the guardian god of prophecy itself), and there are no known cases of the god voluntarily giving a wrong answer or of the god making *bona fide* errors. Above all, and this is the most relevant fact, neither the officials of the sanctuary nor the recipients of the oracles contemplated the possibility that a divine prediction could be wrong at its source (cf. §4.4 and §4.9). This premise, therefore, implies that there were no mechanisms for penalties in the event of a false prediction; such a notion was inconceivable according to this view.

The language of the oracle, however, was dense, often brief, and therefore required interpretation; even the poetic form of many responses contributed to this mysterious allusiveness of expression (cf. §4.1). Thus, in the process of "decoding" the oracular message, there could be a margin of error, but this lay at the level of reception and understanding of the message (on the part of humans) and was not attributed to the source (*i.e.* Apollo: see also Crahay 1974). This insight can be read, for example, between the lines of the chorus at the beginning of Sophocles' *Oedipus Rex* (here referring to Tiresias' role as diviner: cf. *Soph. OT*, 497-506), while examples of oracles that were not easy to interpret are provided, for instance, by Pausanias in relation to the wars between the Spartans and the Messenians (cf. Paus. IV 12 [1-10]; Parke and Wormell 1956, Vol. II, pp. 146–148, nos. 363–365; Fontenrose 1978, pp. 273–274, Q15–Q17).

The nature of Apollinean communication was vividly outlined in a famous saying by Heraclitus (6th–5th century BCE), quoted once again by Plutarch, with commentary: «I imagine that you are familiar with the saying found in Heraclitus [*VS* B 93 Diels and Kranz] to the effect that the Lord whose prophetic shrine is at Delphi neither tells nor conceals [*oute legei oute kryptei*, οὔτε λέγει οὔτε κρύπτει], but indicates [*alla semainei*, ἀλλὰ σημαίνει]» (Plut. *de Pythiae orac.* 404de; transl. F. Cole Rabbitt, London – Cambridge [MA] 1936). This was a widespread perception: in Aeschylus' *Agamemnon* (458 BCE), another prophetess of Apollo, Cassandra, though condemned to be disbelieved even when she spoke truth, clearly states that the oracles of Pytho were expressed in Greek but were nevertheless "difficult to understand" (*dysmathé*, δυσμαθῆ, v. 1255). Her words are preceded by a mention of Apollo with the epithet "*Loxias*" (Λοξίας, v. 1208, repeated by the chorus in v. 1211), which also alludes to the "obliquity", *i.e.* the non-linearity, of the god's expression in his oracular guise (cf. Plut. *de garrulitate* 511b; and Parke and Wormell 1956, Vol. I, p. 40 again).

An example of this is also found in an entry in the Byzantine lexicon *Suda*, which comments on the epithet *Loxias*, offering multiple explanations: «[Another name for] Apollo, he who sends out an oblique voice; for he used to issue oracles obliquely; [*e.g.*] "by crossing the Halys Croesus will destroy a great empire". Or he who makes an oblique journey. For he is the same as the sun» (*Suda s.v.* λ 673, ed. A. Adler, Stuttgart 1933; transl. C. Roth, from "The Suda on Line" project ["SOL"]). Besides the last examples, which are quite clear, the first example ("by crossing the Halys…") refers to a famous episode described by Herodotus (5th century BCE) concerning a consultation that Croesus, king of Lydia, is said to have made at Delphi (after testing the reliability of the oracle, cf. §4.8), to ask whether it would be appropriate to wage war on the Persians. The Pythia, however, did not answer the question directly (with a yes or no) but declared that Croesus "would destroy a great army" if he waged this war (cf. Hdt. I 53; Parke and Wormell 1956, Vol. II, p. 24, no. 53; Fontenrose 1978, p. 302, Q100). Croesus believed that the empire destroyed would be that of the Persians (as a result of his victory), and not his own (as a result of their victory over him: Hdt. I 91, 4). In this way he alone erred (cf. also Hdt. I 66, for a typologically similar response, with Crahay 1974, p. 211: «Delphes annonce aux Spartiates qu'ils arpenteront les terres de Tégée, ce qu'ils feront effectivement, mais en qualité de prisonniers, forcés de cultiver les terres des vainqueurs»). This was therefore a paradigmatic case in which the obliquity or ambiguity of the Delphic oracle's diction was evident, but which also highlighted the possibility of human misinterpretation (cf. also Catenacci 2001, pp. 159 ff.; Giuliani 2000).

In the assessment of human error, a complementary case is provided by the account of the founding of Cyrene, also preserved in Herodotus (IV 150–153; cf. Parke and Wormell 1956, Vol. II, pp. 17–18, nos. 37–38; Fontenrose 1978, pp. 120–123). Grin(n)us, king of the island of Thera (Santorini), had gone to Delphi to seek answers to some questions, but the Pythia avoided answering them and instead instructed him to found a town in Libya. Grin(n)us, however, was old and did not feel capable of carrying out this mission; as a result, Thera was struck by a seven-year drought, at the end of which the inhabitants of the island returned to the Pythia, asked what they should do to overcome the drought, and received a rebuke together with a renewed command to go and found a city in Libya. This, finally, and despite various misadventures, they did.

### 4.3 Authenticity

Also as a consequence of what has already been said, an oracle of Apollo was always "authentic", meaning that it adhered to reality or was effectively predictive. Therefore, ancient anecdotes do not discuss the verification of an oracle's authenticity but, quite often, simply report the fulfilment of an oracle. This could occur at very different times, very soon after the delivery of the prophecy or only much later, and verification usually proceeded from the subjects involved, or from people connected with them. Thus, as an adult, Oedipus fulfilled and "verified" the oracle given to his father Laius, many years earlier (when Oedipus was born), which predicted that he would kill his father and marry his mother; similarly, the descendants of Heracles returned to the Peloponnese only in the third generation after Heracles himself, exactly as the Pythia had predicted when she told Illus (son of Heracles) that this would happen "at the third harvest". For both stories, see *e.g.* [Apollod.] *Bibl.* respectively III 5, 7–9 (48–56); II 8, 2-3 (171–176).

In this regard, it should be noted that the concept of "authenticity" of oracles adopted by Parke and Wormell (and, in a different way, by Fontenrose) differs from that employed in the present study (cf. *e.g.* Parke and Wormell 1956, Vol. II, pp. xxi ff.; Fontenrose 1978, pp. 7 ff.). Those scholars aimed to distinguish, among the mass of documented oracles, those that were actually pronounced on the relevant historical occasions (considered "authentic") from those that were created afterwards (*post eventum*) to legitimise an event with Delphic approval ("inauthentic"). This is, however, a historical-philological distinction based on the evaluation of each case in its historical context, and such evaluation is inevitably subjective. In the ancient sources, we do not usually find an *ante eventum* or *post eventum* distinction, nor suspicion of such; even in narratives of oracles that were very probably invented *post eventum*, for propaganda or similar purposes (on these topics, more extensively, see. also Defradas 1972), the criteria for constructing anecdotes are consistent with standard consultations, as is the notion of the oracle's effectiveness. Thus, the modern concept of "authenticity" does not apply to the verification of the data received from the oracle, but to the possible historicity of the consultation (cf. also Maurizio 1997, esp. pp. 308–312).

However, two issues are linked to the theme of authenticity, as adherence to reality or predictive effectiveness, which will be discussed shortly: the verification of the integrity of the message (cf. §4.4) and, above all, the risk of manipulation of oracles (cf. §4.9). These phenomena, too, belonged to the "human" level of use of the oracles or of their possible alteration: thus, they represented risks inherent in the procedures but did not affect confidence in the validity of the divine message that lay at the basis of the responses.

### 4.4 Integrity

There were no processes for verifying the accuracy of the god's message, *i.e.* that it had not been altered by the Pythia or the priests: these first links in the chain of transmission of the divine message were generally considered authoritative. Documented cases of manipulation of responses, or of the creation of false responses (cf. §4.9), most likely depend on the survival of information regarding the contexts in which the responses were requested and delivered (thanks, for example, to the memories of participants), rather than on the existence of written documents that would have made falsification possible to assess.

After all, the god did not "sign" his responses, he left no mark on them, especially since they were oral in form (cf. also §4.11). To some extent, the solemnity and dense ambiguity of the responses were the only guarantee, or confirmation, of their divine origin. For those who received an answer from a priest or directly from the Pythia, therefore, there was no means of ensuring that this message was exactly the one inspired by Apollo (cf. §4.1): it was ultimately a matter of faith in the god and in his intermediaries (cf. also Dodds 1951, pp. 119–120). Indicative of the prestige of the priests, at least as conventionally claimed, is the invocation with which Pindar's *Paean* VII opens, composed for the inhabitants of Delphi on the occasion of a city festival and addressed precisely to the «glorious prophets/soothsayers of Apollo» (cf. Pind. fr. 52i Mahler, v. 1).

A very curious case of safeguarding the oracular message and limiting the risk of altering the answer (and perhaps even the question) is documented by an important inscription: cf. *IG* II$^2$ 204, with Parke and Wormell 1956, Vol. II, pp. 106-107, nr. 262; Fontenrose (1978), p. 251, H21; Amandry (1997), pp. 202-204; De Martinis (2017). Around 350 BCE, the Athenians asked the god of Delphi what to do with part of the consecrated land of the sanctuary of Eleusis, namely, whether to rent it out or leave it as it was. The two options were formulated as questions, transcribed on tin plates, rolled up, sealed, and placed in a vase so that they could not be distinguished. Then, in a public ceremony, one question was randomly drawn from the vase and placed in another golden vase (*hydria*, a rather large vessel used to collect and transport water), and the other was drawn and placed in a silver one; finally, both vases were sealed. Representatives of the city were then sent to Delphi and asked the god how the Athenians should act, whether according to the instructions in the golden jar or those in the silver one. Thus, the two jars were not even to be opened in Delphi: the god was expected to indicate which jar contained the instructions to be followed, that is, the (supposedly) best course of action for the city. The chosen vase would then be opened only in Athens, again in the presence of the citizens, according to the god's decision.

### 4.5 Availability

According to Plutarch, the oracle was originally "operating" only once a year (specifically on the seventh day of the month of Bysios, a day associated with the birth of Apollo), but later became available for consultation once a month (with the restrictions we will discuss shortly), probably always on the seventh day of the month: cf. Plut. *Quaest. Gr.* 9, 292ef (with *de Pythiae orac.* 398a, and the allusion of Eur. *Ion* 417-421); Amandry (1950), pp. 81-85; Burkert (1983), pp. 116 ff.

These indications, however, should refer to the actual oracular consultations of the Pythia, involving the formulation of a question and, above all, the delivery of a more or less detailed response. In practice, it is possible that "minor responses" (simple questions requiring only a short answer, such as "yes" or "no" to questions like "is it better to do this or that?") were also provided on other days. For these responses, moreover, it cannot be excluded that the Pythia (*i.e.* divination by inspiration) was not employed, but that procedures similar to drawing lots were adopted. The documentation on this subject is very scarce and ambiguous, as well as relatively late (perhaps not earlier than the 4th century BCE). It is possible that pebbles of two colours, or beans, were used, which signified a positive or negative response depending on their colour (cf. Parke and Wormell 1956, Vol. I, pp. 18–19).

It should also be noted that the Oracle of Delphi was inactive for three months each year, when Apollo was believed to leave Delphi to visit the Hyperboreans. During these months, the sanctuary was perhaps ruled by Dionysus (this is, at least, what can be inferred from later sources such as Plutarch, *de E apud D.* 388e–389c; cf. also Parke and Wormell 1956, Vol. I, pp. 11–13), although this god did not deliver responses through the Pythia. It is possible, however, that minor answers continued to be given by the sanctuary during this period.

To better understand the availability of the Oracle, an interesting case is reported by Herodotus and concerns a consultation that the Athenians are said to have carried out during the Persian Wars to learn what they should do (evidently during the final stages of the conflict, and certainly before the Battle of Salamis in 480 BCE): cf. Hdt. VII 140–143 (with Parke and Wormell 1956, Vol. I, pp. 169–171; Vol. II, pp. 41–42, nos. 94–95; Fontenrose 1978, pp. 316–317, Q146–Q147; Maurizio 1997, pp. 329 ff.; Catenacci 2001, pp. 162–164; Bonnechere 2013, esp. p. 81; Mari 2017, pp. 667–668). If this episode is considered historical, we do not know whether it was an exception to the usual procedures (cf. *infra*). According to Herodotus, this was also a case of an anticipated response (cf. §4.6), since the representatives of the city of Athens had not yet formulated their question when the Pythia burst out with a terrible prophecy, which it is appropriate to quote in full to understand how the oracle expressed itself:

«Wretches, why sit ye here? Fly, fly to the ends of creation,

Quitting your homes, and the crags which your city crowns with her circlet.

Neither the head, nor the body is firm in its place, nor at bottom

Firm the feet, nor the hands; nor resteth the middle uninjur'd.

All – all ruined and lost. Since fire, and impetuous Ares,

Speeding along in a Syrian chariot, hastes to destroy her.

Not alone shalt thou suffer; full many the towers he will level,

Many the shrines of the gods he will give to a fiery destruction.

Even now they stand with dark sweat horribly dripping,

Trembling and quaking for fear; and lo! from the high roofs trickleth

Black blood, sign prophetic of hard distresses impending.

Get ye away from the temple; and brood on the ills that await ye!».

[Transsl. G. Rawlinson, London – New York 1910]

This response obviously caused the Athenians to fall into despair, but an eminent man from Delphi advised them to try again, this time presenting themselves as supplicants to the god (*i.e.* to the Pythia) and holding olive branches in their hands. The Pythia then gave a second response, which Herodotus also recorded:

«Pallas has not been able to soften the lord of Olympus,

Though she has often prayed him, and urged him with excellent counsel,

Yet once more I address thee in words than adamant firmer.

When the foe shall have taken whatever the limit of Cecrops

Holds within it, and all which divine Cithaeron shelters,

Then far-seeing Jove grants this to the prayers of Athene;

Safe shall <u>the wooden wall</u> continue for thee and thy children.

Wait not the tramp of the horse, nor the footmen mightily moving

Over the land, but turn your back to the foe, and retire ye.

Yet shall a day arrive when ye shall meet him in battle.

Holy Salamis, thou shalt destroy tbe offspring of women,

When men scatter the seed, or when they gather the harvest».

[Transl. G. Rawlinson, London – New York 1910]

At this point, a great debate arose in Athens regarding the interpretation of the "wooden wall", which was supposed to be the only barrier against the enemies' advance. It is interesting to note that, according to Herodotus, those involved in the debate were primarily the elders of the city and the interpreters of dreams: two categories that could be considered particularly authoritative in understanding the oracular message, respectively for their life experience and their professional expertise. Ultimately, however, it was Themistocles who tipped the balance in favour of interpreting the wooden wall as a fleet of ships to be used against the enemy, and this makes it clear that the interpretation of an oracle could be influenced by the authority of those advancing one explanation over others (cf. also §4.10).

From the point of view of the consultation method, however, the case is significant for other reasons. Double consultations of this type must have occurred very close together and, if not on the same day, at least within a few days. If this were the case, however, it would raise the question of the Pythia's availability on more than one day per month, for which an explanation would have to be found, unless we attribute it to the exceptional circumstances of the war that had affected nearly all Greek cities. Another consideration concerns the nature of the two responses, and thus relates more to the accountability and authenticity of the Oracle (cf. §4.2 and §4.3). In a very short time, the Pythia produced two rather different responses. On closer inspection, however, the information provided diverges only in perspective: the first oracle focused on the destruction of Athens (which indeed occurred) and, with its narrow focus, could only generate discouragement among the Athenians; the second broadened the view (also the geographical one), including indications of the place that would mark the Greeks' victory (Salamis) and the means by which it would be achieved (the fleet). In essence, the god could not be accused of error or self-correction: it was the new consultation, carried out in a supplicatory spirit, that led Apollo to provide a broader vision and an important glimmer of hope for Athens. On cases of double consultation of the same oracle, cf. also Bonnechere (2013) (with conclusions such as this: «il ressort que la double consultation, pour le consultant privé ou public, consiste seulement à préciser la réponse de l'oracle qu'il estime trop vague» [p. 93]).

### 4.6 Latency

The Pythia is usually depicted, especially in myths, as able to respond at once to the question she was asked.

Plutarch, in fact, reports that the answer could be given even before the question was formulated; to explain this, he invokes the god's introspective power, capable of «understanding the mute and hearing the voice of those who did not speak» (*Plut. de garrulitate* 512e). The divine nature of the event, in any case, did not entail any interval for processing the data or reflecting on the question: the god's omniscience meant that an answer was available immediately after the question was asked (if not before, as already noted; cf. §4.5). We do not know, however, whether this was always the case in reality, and indeed some scholars have hypothesised that the latency times could have been longer: the request could have been submitted to the priests, perhaps transcribed, and then transmitted to the Pythia; the response, in turn, could have required a similar process. In both cases, the timing would not be quantifiable, at least for us.

### 4.7 Accessibility

As early as the Archaic period, the Oracle of Delphi was available to all Greeks (not only those from Phocis) and even to non-Greeks: a prime example is the case of Croesus, king of Lydia (cf. §4.2). Although there were no direct restrictions preventing access to the oracle, the nature of the procedures probably involved (1) a financial commitment on the part of the applicants and (2) the manifestation of the god's consent to express himself through the Pythia.

(1) The financial commitment derived from the fact that each applicant, once purified, made an offering for the service they would receive; then, after being admitted to the consultation, the applicant had to sacrifice a medium-sized animal (usually a sheep or a goat). The offering and the sacrifice, therefore, required financial resources. There was also an order for consulting the oracle, at least in Classical times: «The city of Delphi and its citizens had the first right to consult it, and after them those cities or individuals to whom, as a sign offriendship, the city of Delphi had granted the privilege of *promanteia*, that is, the right to consult the oracle on the same terms as Delphi» (Bowden 2005, p. 17). However, if there were numerous applicants, it is possible that a draw would have been used (cf. *e.g.* Aesch. *Eum.* 31–32).

(2.) The god's favour for the consultation was itself a prerequisite that had to be fulfilled. Before the sacrifice, it was necessary for the animal victim not only to move its head, as if to show acquiescence to the sacrifice, as was often expected elsewhere, but also to be seized by spasms. This was the only sign of the god's presence: the victim's convulsions served as a sort of prelude to those that would later seize the Pythia, when she became possessed by the god, and as a guarantee that her responses truly came from him (cf. Plut. *de def. orac.* 435bc; cf. also §4.1 on Pythia's inspiration). This procedure was considered essential and could not be forced, at the risk of provoking the wrath of the god: Plutarch recounts an episode in which the priests compelled a sacrificial animal to react by administering repeated and excessive libations, thereby violating normal practice; the Pythia then produced a strange response and was overcome by such distress that she ran away screaming. Plutarch notes that she recovered, but survived only seven days after this episode: in his view, this was an example of Apollo's punishment of those who had forced the normal oracular procedure (cf Plut. *de def. orac.* 438ab).

However, it is not impossible that access to the Oracle was organised differently or varied according to circumstances and periods. For instance, we have already considered the example in Herodotus (VII 140–143: cf. §4.5) regarding the Athenians' two closely successive consultations of the Oracle. A different entry procedure is documented in the episode involving Calonda (also known as "Raven"), who murdered the poet Archilochus. Calonda went to the Oracle to request a response but was initially rejected (perhaps twice) because he was tainted by the impurity of murder. He then received the suggestion (or order) to purify himself of the crime he had committed, perhaps by going to a specific place (Cape Tainaron, according to some sources). This purification would have put an end to Apollo's wrath towards him, and we may therefore deduce that he subsequently regained access to the Oracle (cf. Heraclid. Lemb. Περὶ πολιτείων, Fr. 25 Dilts [Παρίων]; Plut. *de sera num. vind.* 17, 560de; *Suda* s.v. α 4112 Adler; Parke and Wormell 1956, Vol. I, p. 397; Vol. II, pp. 3–5, nos. 4–5; Fontenrose 1978, pp. 242 e 287, Q58).

### 4.8 Centralization

Since the request for an oracular response concerned the divine sphere and did not simply focus on human practices or knowledge, there was no question about the "centralisation" of data, *i.e.* the source of the responses. On the contrary, the divine omniscience of Apollo (or of any other deity who expressed himself through an oracle) guaranteed the authenticity of the information provided in response, so it was not subject to scepticism (cf. also §4.3). Nor did the reception of an unsatisfactory or obscure response from an oracle necessarily imply a new request for another response from a different oracle, whether of the same god or of another deity, for confirmation. A similar action is documented by Herodotus, but significantly it was carried out by a non-Greek, *i.e.,* by someone outside the religious beliefs of the Greek world, and therefore outside the system of trust in oracular authority. This is the famous episode of the testing of the oracles devised by Croesus (cf. Hdt. I 46–49; Parke and Wormell 1956, Vol. I, pp. 129–130; Vol. II, pp. 23–24, no. 52; Fontenrose 1978, pp. 301–302, Q99; on Croesus and Delphi, cf. also Crahay 1974, pp. 215–219).

According to this anecdote, the king of Lydia, concerned about the growing power of the Persians and uncertain whether to start a war against them (cf. §4.2), decided to test the effectiveness of various Greek and Libyan shrines and then to pose his question to the oracle that proved the most reliable. Croesus employed the following method of verification: he sent his emissaries to the main prophetic centres and ordered them to ask each oracle, on the hundredth day after their departure from his court (thus simultaneously and at a predetermined time), what Croesus himself was doing at that precise moment. The emissaries would then transcribe the answers and report them to him. At this point, Herodotus focuses his attention on Delphi: «the moment that the Lydians entered the sanctuary, and before they put their questions the Pythia thus answered them in hexameter verse: "I can count the sands, and I can measure the ocean; / I have ears for the silent, and know what the dumb man meaneth; / Lo! On my sense there striketh the smell of a shell-covered tortoise, / boiling now on a fire, with the flesh of a lamb, in a cauldron. / Brass is the vessel below, and brass the cover above it"» (Hdt. I 47; transl. G. Rawlinson, London – New York 1910, adapted). When all the emissaries returned to Croesus and brought back the transcriptions of the responses, the king declared that only the oracle of Delphi had proven reliable: on the hundredth day, indeed, Croesus had taken a lamb and a turtle and boiled them in a bronze cauldron, just as the Pythia had revealed (cf. Hdt. I 48).

Herodotus, incidentally, does not specify what erroneous answers the other oracles gave (cf. Hdt. I 47, 2): the story is meant above all to exalt and defend Delphi, though the historian is careful to avoid direct criticism of other shrines. Croesus, for his part, would later demonstrate his veneration for the Delphic sanctuary with magnificent gifts; yet he would also become an exemplary case of failing to interpret correctly the oracle's response (cf. §4.2), perhaps through haste and neglecting to consult experts who could have warned him about the ambiguities of Pythian diction.

### 4.9 Collusion

The problem of potential manipulation of oracles remains open and, as of today, has no definitive solution, mainly due to gaps in our documentation. However, it is worth distinguishing between two aspects of the issue.

First of all (1.), we do not know the true nature of the procedures that constituted the oracular act, nor, above all, the attitude or degree of good faith that motivated the officers of the sanctuary. And when we speak of "good faith", we must assume a genuine adherence to, and participation in, the belief that what occurred during an oracular response was truly directed by the god, whatever the underlying phenomenon may have been, something we cannot clearly define. Various hypotheses have been formulated: it may have been a state of psychic suggestion, alteration, hypnosis, or semi-consciousness that produced messages subsequently interpreted as the voice of Apollo. The Pythia, therefore, may have been acting in complete good faith, or she may have been influenced by personal interests in formulating certain responses, thereby shaping them. The same could be true of the priests (as many interpreters are more inclined to believe) or of anyone else involved in the chain transmitting the god's message to the enquirer. Some reflections by Parke and Wormell remain pertinent in this regard: «the confused and disjointed remarks of a hypnotised woman must have needed considerable exercise of imagination to reduce them to the form of a response. In this must have lain the chief temptation for the priests. Where must they draw the line and confess that they were merely reading their own thoughts into the Pythia's gabble? No doubt in this matter individual priests on particular occasions took a very different course» (Parke and Wormell 1956, Vol. I, p. 39 [and, more in general, cf. pp. 36 ff.]; cf. also Ustinova (2021). The perception of this risk, however, must already have been ancient, or at least recognisable in certain accounts from Classical Athens. It can be read between the lines, for example, in the words that Sophocles places in Jocasta's mouth when she recalls the oracle received by her husband Laius: «An oracle once came to Laius, I will not say from Phoebus himself, but from his servants …» (Soph. OT 711–712; transl. H. Lloyd-Jones, Cambridge [MA] – London 1994; cf. also Parke and Wormell 1956, Vol. I, pp. 192–193 [n. 27]). This suggests that, behind the words received from the oracle, one might suspect the intervention of the temple officiants rather than the simple, direct utterance of the god himself. Similarly, the case of the sealed metal jars in Athens (cf. §4.4) could also be interpreted as an attempt by the Athenians to protect their request from interference outside of the divine realm.

Second (2), if we look beyond the actions performed within the sanctuary during oracular procedures, ancient sources from at least the Archaic and Classical periods consistently reveal substantial trust in oracles and great respect for their responses among those who consulted them. This attitude was probably widespread (certainly not unanimous, but at least widely shared: cf. also Lombardo 1972; Parker 1985) and may have begun to weaken only from the Hellenistic period onwards, or slightly earlier, with the emergence of philosophies more open to relativism or scepticism in religious matters. That said, even in the Archaic and Classical periods, there are documented cases of manipulation of oracles. Two of them, both reported by Herodotus, are particularly significant.

The first dates back to the second half of the 6th century BCE: the Alcmaeonids, exiled from Athens by the Pisistratids (the city's tyrants), managed to bribe the Pythia so that she would tell any Spartan who consulted the oracle that it was imperative to liberate Athens from the Pisistratids. The Spartans, persuaded by this divine injunction, undertook the campaign but suffered a heavy defeat (cf. Hdt. V 63; Parke and Wormell 1956, Vol. II, pp. 35–36, no. 79 [and also Vol. I, pp. 145–147]; Fontenrose 1978, pp. 309–310, Q124).

The second example concerns events that took place a few decades later: Demaratus, king of Sparta, was deposed after the Pythia, convinced by an accomplice of Demaratus' political rivals, pronounced that Demaratus himself was not the son of the previous king (Ariston). As a result, Demaratus was delegitimised. Herodotus, however, also reports the fate of the Pythia: when the conspiracy was uncovered, she «ceased to exercise her role» (ἐπαύσθη τῆς τιμῆς), or, more likely, was removed from her position (cf. Hdt. VI 66; Parke and Wormell 1956, Vol. II, p. 38, no. 87 [and also Vol. I, pp. 161–162]; Fontenrose 1978, p. 314, Q137). Furthermore, the historian also recounts the madness and violent death of Cleomenes, one of Demaratus' enemies and the instigator of the oracle's corruption, adding that this outcome was considered by most Greeks to be divine punishment for having manipulated the Pythia (cf. Hdt. VI 75; cf. also Jacquemin 1995; Giuliani 1998). Even these cases, however, did not undermine general confidence in the oracle (cf. also Lloyd-Jones 1976, pp. 67–68). On the contrary, the punishment of those who manipulated or forced the oracular process was seen as confirmation of the Oracle's sanctity, as well and as a sort of deterrent, a divine warning not to transgress the god's will.

### 4.10 Lazy Equilibrium

As oracular responses were often difficult to interpret, recipients, whether individuals or groups, could turn to experts for help in evaluating the messages. These experts were usually familiar with the language of oracles and with the Pythia's enigmatic mode of expression (cf. §4.2), and they placed their expertise at the service of those seeking a less improvised reading of a response. This category also included prophets or soothsayers, who were believed to be capable of achieving greater harmony with the divine spirit and thus of correctly interpreting the meaning of Apollo's words. That said, even prophets were sometimes unable to discern the ambiguities of Apollo's diction. A case in point is that of the Messenian *manteis* ("diviners"), who struggled to interpret an oracle (cf. Paus. IV 12, 4; and §4.2 again).

On the other hand, a simple way of reducing the risk of error was to share the oracular response with the entire community, enabling it to be interpreted collectively and subjected to scrutiny by multiple minds. This collective interpretation reduced the risk of individual misjudgement. An example of this is the double oracle given to the Athenians before the Battle of Salamis, which we have already discussed (cf. §4.5).

### 4.11 Freeloading

Our sources confirm instances of responses in which entire verses appear repeated. These typically contain powerful images that are ambiguous and multifaceted, and could therefore be applied to a variety of situations (cf. also §4.2). This repetitive character, however, did not result from the laziness of the issuers (Pythia/Apollo) or from any standardisation of messages, but rather from the very nature of the oracles, which were expressed in traditional language and thus were intrinsically open to repetition. In the oracles, the same "formulas", expressions, and images were reused in different contexts to convey similar ideas, but were always adapted to suit the specific circumstances of each person seeking advice (cf. also Parke and Wormell 1956, Vol. I, pp. 280–281; Vol. II, pp. xxx–xxxi).

Already famous in antiquity, and even proverbial, was a response to the question of who were the strongest among the Greeks. According to some sources, the questioners were the inhabitants of Aigion (in Achaia); according to others, they were the people of Megara. In both versions, after listing a series of excellences (the most fertile land, the fastest horses, etc.), the Pythia replied in hexameters: «But you, inhabitants of ..., are neither third nor fourth nor twelfth in consideration or account», thereby extinguishing any hope of supremacy on the part of the enquirers. The accounts of this oracle are numerous and complex (cf. Parke and Wormell 1956, Vol. II, pp. 1–2, no. 1; Fontenrose 1928, pp. 276–278, Q26; Bühler 1982, pp. 270–276, no. 35; Maurizio 1997, pp. 323–326), and it is probably impossible to establish which version of the story (and the response) was the original (but cf. Parke and Wormell 1956, Vol. I, pp. 82–83, for a different view). On the other hand, it is not at all unlikely that a similar question, posed by two different enquirers, would have received a similar response, based precisely on the reuse of traditional language. In other words, and more broadly, the formulaic nature of oracular diction did not indicate a repetitive or unimaginative attitude on the part of the speaker, but was simply a defining feature of oracular language. This reflected its social and cultural dimension (cf. also Amandry 1997, pp. 204 ff.; Maurizio 1997, pp. 312 ff.).

### 4.12 Queries

As explained, historical and quasi-historical queries were first classified according to Bartholic (2022), following the rationale summarized in Table 1. The Computational Queries category is omitted, as no responses in the corpus corresponded to this type.

Table 1. Author Rationale for classifying Delphic queries according to Bartholic (2022)

| Category | Who *could* answer (in principle)? | Nature of knowledge sought | How the "truth" of the answer is established | Example question | Analogy in blockchain oracles |
|---|---|---|---|---|---|
| Discernible Event | Many observers (public, empirical) | Observable fact or outcome | Broad agreement through direct observation | "Will the sun rise tomorrow?", "Has the race been won?" | Data feeds from sensors or public APIs (weather, price, sports results). |
| Sanctioned Event | Restricted subset of *authorities* (priests, judges, the god as ultimate arbiter) | Normative or moral judgment; permission, legitimacy | Authoritative declaration by a recognized source | "Who should lead?", "Is it pious to refuse the truce?" | Governance oracle or validator vote deciding legitimacy or upgrades. |
| Recondite Event | Only one exclusive source (the god) | Hidden factual knowledge, past, future, or concealed state | Revelation verified "after" the fact, if at all | "Where are Orestes' bones?" | Oracle accesses off-chain data unknown to users (e.g., hidden data). |

| | | | | | | |
|---|---|---|---|---|---|---|
| Ambiguous Event | Multiple honest interpreters could answer "differently" using the same data | Mixed or indeterminate meaning; semantic or contextual conflict | No single authoritative truth; interpretation decides outcome | "Where shall I settle?" | Conflicting oracle data feeds, equivocal signals requiring governance resolution. | |
| Non-Event | No one can truly answer. Question has no factual or normative grounding | Open-ended, metaphysical, or rhetorical inquiry | Not verifiable or falsifiable; generates reflection, not decision | "What is best for man?", "Why are gods unjust?" | Questions outside the oracle scope (philosophical, non-computable prompts). | |

*Author elaboration

Following classification, a lexical analysis of the oracle answers was conducted using Python-based computational tools. The aggregated results, expressed as category averages, are reported in Table 2.

Table 2. Lexical analysis over Delphic queries

| Query Type | Occurrences | Avg. Word Count | Avg. Shannon Entropy | Avg. Modal Density | Avg. Hedge Density | Avg. Polarity | Avg. Subjectivity |
|---|---|---|---|---|---|---|---|
| Ambiguous | 15 | 16.93 | 3.55 | 0.029 | 0.00416 | 0.24 | 0.32 |
| Discernible | 58 | 23.13 | 3.94 | 0.067 | 0.00115 | 0.1 | 0.22 |
| Non-Event | 21 | 30.36 | 4.12 | 0.041 | 0.00242 | 0.25 | 0.32 |
| Recondite | 20 | 32.4 | 4.21 | 0.045 | 0.00039 | 0.21 | 0.45 |
| Sanctioned | 53 | 27.33 | 3.98 | 0.043 | 0.00166 | 0.09 | 0.26 |

Results show that Recondite questions are associated with the highest average word count (32.4) and Shannon entropy (4.21), indicating that verbosity and lexical complexity increase when the subject concerns matters inaccessible or unknowable to most audiences. Conversely, complexity decreases for questions that can be understood by specialized (3.98) or general (3.94) groups of individuals.

Modal density reveals that authoritative or prescriptive language is most frequently employed in responses to Discernible questions (0.067), while more indeterminate or interpretive formulations characterize Ambiguous questions, for which hedge density reaches its peak (0.00416). Nonetheless, hedge density remains low across all categories, confirming that Delphic pronouncements were generally assertive and authoritative. This supports the view that the oracle's responses were not inherently vague or cryptic as often portrayed in later tradition. Rather, the perception of ambiguity likely derives from legendary or fictional reinterpretations.

Regarding polarity and subjectivity, the most neutral and objective answers occur in Discernible (0.10; 0.22) and *Sanctioned* (0.09; 0.26) questions, where the oracle's tone appears factual and directive. In contrast, Recondite questions produce the highest subjectivity (0.45), reflecting the epistemic inaccessibility of matters answerable only by the god itself.

Overall, the analysis suggests that the Delphic oracle performed most effectively when addressing questions of public relevance or those confined to a well-defined domain of expertise, such as religious or political matters. These were contexts in which the oracle's interpretive framework, institutional authority, and accumulated knowledge could operate within recognized epistemic boundaries. Drawing a parallel with blockchain oracles, the findings imply that an optimal oracle likewise provides more reliable and verifiable outputs when responding to queries situated within its domain of specialization or when dealing with standardized, publicly accessible data. For instance, as a religious institution, the Delphic oracle excelled in providing guidance on ritual practices and ceremonial decisions. Similarly, a blockchain oracle dedicated to price feeds will yield the most accurate and trustworthy results when queried about asset prices, where market data is structured and transparent. Conversely, ambiguous or misleading queries, whether directed to a priestess at Delphi or to a decentralized oracle network, are likely to produce inconsistent or biased outputs regardless of the oracle's intrinsic reliability or sophistication. In both cases, the clarity and framing of the query determine the quality of the answer, reaffirming the structural continuity between ancient divinatory reasoning and contemporary oracle design.

The following tables (Table 3,4,5) summarize the findings retrieved in this section, drawing parallels between the ancient Delphi and modern blockchain oracles. The next section instead discusses these findings in relation to academic and practitioner literature.

Table 3. Delphic consultation procedures and blockchain oracle intuitions.

| Characteristics | Delphic model | Intuitions for Blockchain oracles |
|---|---|---|
| Attributability | The origin of the information was the god Apollo, who was able to read Zeus' mind. His identity is clearly known to the petitioner and indisputably trusted | The data source should always be transparent and known to be reliable. |
| Accountability | The process of communicating divine insights could face challenges, leading to misunderstanding and misinterpretation. It's the responsibility of the petitioner to interpret the response correctly. The Delphic institution itself did not face any formal sanctions, except for a negligible loss in reputation. | The oracle should not be accountable for the quality of data provided. It is the responsibility of the petitioner to select a proper and reliable oracle. |
| Authenticity | Ancients treated oracles mostly as authentic, focusing on their fulfillment over time. Modern scholars assess authenticity between oracles actually pronounced at historical events and those retroactively created to justify human action. | Time should be the parameter used to distinguish reliably authentic oracle data. |
| Integrity | As prophecies were orally transmitted, no signature or mark from Delphi was provided. The petitioner could, on the other hand, formulate the question using specific methods (e.g., marked amphoras) to enhance the integrity and reliability of his request. | The Oracle response mechanics should be standardized. It's the petitioner who may implement additional querying mechanics to enhance the reliability of the process. |
| Availability | The Delphic oracle was available for complex consultations only on designated days. Binary responses for ordinary matters could also be given on other days without involving Pythia. Extraordinary circumstances may also allow consultation outside designated days. | The main Oracle engine should be available only for specific matters under given conditions. Different conditions may be applied for urgent or straightforward matters. |
| Latency | The response time of the Pythia was generally immediate | Oracles should answer questions with the lowest possible latency. |
| Accessibility | The Delphic oracle was, in principle, openly available. However, consultations were not effortless as they required following specific procedures, sacrifices, and donations. Moreover, as categories of people had privileges, others had to patiently wait for their turn. | Oracles should be open, but their services should not be offered for free and effortlessly. It is also plausible to organize responses according to priority lists. |

Table 4. Delphic consultation risks and blockchain oracle intuitions.

| Risk Type | Description | Intuitions for Blockchain oracles |
|---|---|---|
| Centralization | The Delphic oracle was somehow centralized, but this was not considered a problem, as it was a source of Divine Authority. Petitioners could, every now and then, test the reliability of the oracle by querying others about the same questions (e.g., Croesus) | Centralization of oracles should not be perceived as a negative characteristic if the oracle is reliable. Every now and then, it can be tested to confirm its integrity, however. |
| Collusion | Known cases of collusion to manipulate Delphic responses are rare. When the Pythia is discovered, she is permanently deposed from her duty. | When an oracle is spotted manipulating the outcome, it should permanently cease to operate or to be queried. |
| Lazy Equilibrium | Delphic messages could be difficult to understand, even to a specialized audience. Very few people could efficiently interpret them. To reduce the risk of error and accountability, the oracle response was shared with the community, enabling a collective interpretation. | It's unrealistic to hypothesize multiple reliable reporters or interpreters. For complex answers, off-chain open consultation may be leveraged. |
| Freeloading | Ocular responses were standardized in the form of many answers that appear repeated. However, the final response slightly differs according to the specific question. | Oracles should be standardized to increase efficiency while still being capable of offering personalized responses. |

Table 5. Delphic queries and blockchain oracle intuitions

| Query types | Delphic response format | Intuitions for blockchain oracles |
|---|---|---|
| Discernible/Sanctioned | Shorter, modal, and more precise response. | Domain-specific oracles should perform better than all-purpose oracles. Data from the public domain should also be more reliable. |
| Ambiguous/Recondite/non-event | Longer, ambiguous, and blurry responses. | Oracle calls concerning non-public or scarcely available data may be less reliable and faulty. |

5. **Discussion**

The present paragraph discusses Delphic consultation procedures explained in the previous section and speculates whether these could still be relevant and useful for a modern blockchain oracle, also investigating if some are already being used in existing blockchain oracle protocols.

**5.1 Attributability in Delphi Vs blockchain oracles**

The analysis starts with attributability, which refers to the ability to know who the data reporter is. In the Delphic scenario, it was always known that the source of information was the God Apollo, an undisputably trusted Entity. This idea is quite different from what research and industry development in blockchain oracles tries to offer. Blockchain oracles aim to be trustless and decentralized. The identity of reporters is usually not known in light of preserving anonymity of systems, while their reliability is ensured by balancing rewards and punishments through game-theoretical schemes (Pasdar, Lee and Dong, 2021). However, a well-known project in web3 called API3 embodies the rationale of Delphic attributability. API3, in fact, relies on the fact that some entities are trusted and competent in a certain sector, and therefore, they represent the most appropriate data source for a specific query (Benligiray, Milić and Vänttinen, 2022). Reporters' transparency may greatly reduce manipulation mechanisms such as sybil attacks, while increasing the reliability of feeds by enabling direct accountability for faulty or imprecise reports.

**5.2 Accountability in Delphi Vs blockchain oracles**

Strictly connected to attributability is the accountability of a data source. As previously explained in the Delphic scenario, the oracle was not held accountable if the prediction didn't materialize, since the responsibility of failure was often translated to the petitioner, such as in the case of Croesus for the war against Persia. It was the petitioner's responsibility to query the right oracle for the right reason, to interpret the message correctly, and to take the most righteous action as a consequence. In case of a negative externality as a consequence of an unwanted or misinterpreted oracle prediction, what the oracle suffered was maybe a loss in reputation, so that users could, if they wanted, query another oracle if they were not satisfied. Similarly, in the blockchain oracle scenario, the responsibility for an incorrect data report is usually taken by the web3 protocol, therefore, by the entity that queries the oracle. Past cases of oracle failure, such as the compound incident, the negative outcome was in fact suffered by the protocol or by the final users, eventually (Caldarelli and Ellul, 2021; Werner *et al.*, 2022). As a direct consequence, the protocol may decide to change the oracle, and the final user may change the Web3 protocol. In this sense, the current blockchain oracle scenario is aligned with the Delphic one, although this is not yet an established standard. A very similar Idea, instead of an oracle market based on the reputation, similar to the Delphic system, was developed in the first days of Bitcoin by a protocol called Reality Keys. The idea was that reporters on reality keys vouched for data with their reputation, and users could freely and openly select a data source based on their faith in one specific data provider (Edgar, 2014; Southurst, 2014; Caldarelli, 2023).

**5.3 Authenticity in Delphi Vs blockchain oracles**

Moving on, the parallelism on the concept of data authenticity is quite interesting, as it is conceptually similar between Delphic and modern blockchain oracles. We consider only Delphic petitions about present matters or consultations, as blockchain oracles do not provide predictions over future events (yet). If the Delphic oracle was asked to provide an opinion about a specific matter, its rationale for verifying truthfulness was to wait until the event was fulfilled. In the case of blockchain oracles, when queried about the outcome of a match, the rationale is to wait a certain number of days so that this information is no longer disputed. So in both cases, the parameter to ensure the authenticity of a report is the time. This intuition was emphasized in blockchain oracle protocols like Truthcoin, where, beyond the mathematical threshold for identifying the data to be reported, its authenticity and indisputability were ensured by a specific time threshold (Sztorc, 2015). After Truthcoin, many other protocols, such as Augur or Uma, adopted time as an acceptance threshold for oracle queries; therefore, we may argue that this is quite an established parameter in the blockchain oracle space (Peterson *et al.*, 2015; UMA, 2018).

### 5.4. Integrity in Delphi Vs blockchain oracles

Another interesting aspect to consider is the integrity of Delphic oracle, which involved determining if the message really came from the Pythia or from someone else who impersonated the priestess. Given the fact that oracular messages were transmitted orally, no signatures were provided by the oracle; thus, the only way to be certain of the origin of the message was to interact directly with the Pythia. For those who received the message through a third party, only the solemnity and tone of the message were used to ensure its integrity. It was also possible for the petitioner to implement some additional schemes to make sure the message was not manipulated, for example, using sealed containers to be opened only at the delivery. Ex post, historians debated the authenticity of Pythian messages by anchoring the retrieved writing to verified historical events. The core idea is that the oracle did nothing to prove its identity in the message, and it's the petitioner who has the responsibility of ensuring the message's provenance and its integrity through various means. This procedure closely resembles client-supplied authenticity proof oracles such as DECO or Provable, in which the data source does not sign the message for blockchain consumption; instead, requesters attach a cryptographic transcript that binds content to the origin (D-Nice, 2017; Zhang *et al.*, 2020). Direct contact with the oracle is again observed with providers such as API3 and Ex post integrity via community parallels with optimistic oracles and dispute resolution, in which a value is accepted unless disputed (optimistic) or the truth is settled after the event by anchoring the claim to verifiable public records (dispute resolution)(UMA, 2018; Lesaege, Ast and George, 2019; Tellor, 2020; Benligiray, Milić and Vänttinen, 2022).

Consultation through sealed urns instead requires an extensive discussion. To summarize, on a specific occasion, the Athenians queried the Delphic oracle on how to deal with some lands, and since they didn't wish to influence oracular choice with the question, they decided to seal the two answers in closed urns, a golden and a silver one. As impersonating the god Apollo, the Pythia didn't need to read the content of the urn to select the right one. They sent the sealed urns to Delphi in the presence of three witnesses, one from the council and two from Athens, to confirm the choice of the Pythia, who could not sign the urn. They then returned to Athens with the chosen urn, unsealing and publishing the contents.

Projecting this system in a modern Oracle scenario, it would be organized as follows.

In the "Commit phase", the petitioner posts two commitments H0 = hash(m0||r0) and H1 = hash(m1||r1) on-chain, or to a Decentralized Oracle Network (DON), where "m" is the message and "r" is a random number inserted into the message to make sure the hash is non deterministic (otherwise anyone knowing the messages can distinguish them once hashed). This setting ensures that the message is "hidden" to any external entity once hashed and "binding" as the content can't be changed later.

The commitment should then be either timestamped when submitted on-chain and then co-signed by multiple independent witnesses to obtain a public tamper-proof record of what was committed.

At this point, an oracle should blindly choose the correct instruction, which is unrealistic. The closest option we may have is a representation of the blindfolded goddess of luck, therefore, a randomness oracle. This randomness oracle should then select one of these two commitments, putting a signature "b", enabling an independent and unbiased choice, preventing any manipulation and frontrunning.

In the "reveal" phase the message should then be published so that the petitioner can submit (m_b, r_b) and the contract verify that hash(m_b||r_b) == H_b. If anything were altered, verification would fail.

Organized like this, it's a random commit reveal oracle with witnesses. It's similar to Chainlink VRF (Breidenbach *et al.*, 2021), but with some additions that may be introduced when randomness must be strictly ensured (e.g., high-stakes lotteries).

### 5.5. Availability in Delphi Vs blockchain oracles

The availability of the Delphic oracle is very fascinating as it resembles some schemes of modern blockchain oracles. As explained, the Pythia that could answer complex questions had limited availability (once per month), while for simple yes/no questions, priests were available more often. The case of UMA oracle, for example, is very similar, as it offers an optimistic oracle response for simple questions and a more complex and longer response mechanism for more difficult or delicate matters (UMA, 2018). Therefore, in line with Delphic procedures, it is certainly plausible to have two response mechanisms depending on the importance and/or difficulty of the query in which the most complex one has less availability. The most complex and secure one is expected to be less available than the simpler one.

**5.6. Latency in Delphi Vs blockchain oracles**

As for latency, we observe that the Delphic response was generally immediate. Modern blockchain oracles strive for immediate feed, but as explained above, for complex matters, they prefer a programmed delay to better counter manipulation and ensure data accuracy. So, for data feed, it makes sense to pursue a low level of latency, while for more complex matters, it is better to maintain a higher level of latency, unlike Delphic design.

**5.7. Accessibility in Delphi Vs blockchain oracles**

Accessibility is also an interesting matter to compare. Although the Delphic oracle was freely accessible, petitioners from Delphi and some with special rights had priority to query the oracle. Modern oracle networks mirror this setting. While data are publicly verifiable, access to query interfaces or low-latency feeds is often tiered through staking, whitelisting, or paid subscription models. For instance, Chainlink's premium feeds restrict the freshest or low-latency data to subscribing protocols, or API3's Airnode framework for example limits oracle calls to whitelisted smart contracts registered with the data provider. These layered access models reproduce, in digital form, the Delphic differentiation between ordinary petitioners and those granted priority consultation rights (Breidenbach *et al.*, 2021; Benligiray, Milić and Vänttinen, 2022).

Concerning payment, this aspect was also peculiar at Delphi. The temple received offerings for divination, but these were formally donations to a sacred institution providing a communal service rather than commercial fees. In a similar vein, blockchain oracles can be viewed as providers of a public good, data integrity, and reliability for decentralized systems. Many Web3 oracle networks are operated by foundations, and under appropriate regulation, accepting voluntary donations could represent a sustainable model, particularly when certain baseline services are offered free of charge. Such an approach would preserve the oracle's public-service ethos while ensuring financial support without compromising neutrality.

**5.8. Consultation Risk in Delphi Vs blockchain oracles**

Concerning a comparison among risks in consulting the oracle, the first to be considered was the risk of Centralization. As explained, the oracle of Delphi was highly centralized, but this was not seen as an issue for Apollo being an undisputed source of truth; however, when skepticism arises, as in the case of Croesus, decentralization could be leveraged to evaluate the reliability of the oracle by querying multiple oracles and verifying the reliability of their answer.

For modern blockchain oracles, instead, centralization is seen as a limit and something to prevent. Therefore, projects such as Chainlink, Pyth, DIA, or Band query multiple data sources in order to answer a query, and finally accept a value that stays inside a threshold (Zhao *et al.*, 2022). However, this mechanism makes the oracle activity slow, complex, and costly, negatively impacting interoperability, accessibility, and scalability. A system similar to the Delphic one that has a centralized data source and, every now and then, leverages multiple data sources to test its reliability would greatly reduce costs and complexity.

As for collusion and bribery, Delphic history contemplated these types of circumstances. But the rare event of manipulation was followed by the certainty of punishment. The punishment at the time of Delphi was to resign from the role of god emissary. In a modern Oracle scenario, a similar mechanism can be implemented that removes the node or data source permanently in case of manipulation. It's definitely more drastic than current schemes that require a fine or a slash, but it is indeed effective, since if we consider a transparent and known data source, there is no possibility for the same entity to create another account and serve again as an oracle. Being a transparent data source also reduces the chance of a Sybil attack (Douceur, 2002).

Cases of freeloading and Lazy equilibrium instead are hardly encountered in ancient oracle schemes since the oracle was centralized and not voting-based. In Delphi prediction, we encounter cases of standardization of procedure that make responses mostly similar to each other and help their interpretation. In fact, more difficult and unique responses could only be interpreted by experts, and in some difficult cases, interpretation was submitted to public judgment. Today, interpretation can be viewed as the ability of a smart contract to digest data collected from multiple oracles that, if in a different format, would require some adaptation. In this sense, as suggested in multiple academic studies, more standardization in blockchain oracles would be beneficial (Caldarelli and Ellul, 2021).

**5.9. Queries in Delphi Vs blockchain oracles**

The concept of queries is also worth examining. The framework proposed by Bartholic (2022) adopts a more philosophical perspective, which allows for a smoother integration with the analytical approach developed in this paper. In modern contexts, oracle queries generally take the form of calls requesting factual or binary data, most often prices, rather than the complex interpretive questions characteristic of the Delphic oracle. Nevertheless, the intuition that emerges is straightforward. Oracles tend to perform more reliably when addressing data that are either "publicly observable" or "domain-specific". For example, prediction-market oracles such as Augur rely on outcomes that are verifiable by the public, leveraging the "wisdom of the crowd." Similarly, protocols such as Uniswap provide reliable price data precisely because they operate within their area of specialization. Hence, as with the Delphic oracle, reliability increases when the question lies within the oracle's legitimate epistemic domain, either because the truth is collectively observable or because the oracle itself possesses intrinsic expertise in that field.

## 6. Conclusion

This research investigated the consultation procedure of the Delphic oracle to draw parallels with modern blockchain oracles, with the aim of proposing innovation in this domain. By standardizing the classic consultation procedure and leveraging blockchain oracle characteristics, a framework is obtained that can also be used to analyze and classify other classic oracles, thereby expanding research in this domain. This study has limitations due to the scarcity of historical material and the necessary compromises that had to be made to assertively establish Delphic procedures, on which we will never have absolute certainty. Limitations were also present in the analysis of queries, where, on the one hand, despite leveraging prior studies, we had to subjectively interpret their underlying rationale, and on the other hand, the lexical analysis had to be based on the text provided by Fonternose (1978), in which Delphic queries are translated and sometimes interpreted. Despite the inherent limitations, the results were surprisingly interesting as many parallels were observed between Delphic and blockchain oracles. Thanks to these parallels, many characteristics of modern blockchain oracles, such as anonymity, openness, and decentralization, may be revised in light of a more efficient solution. Building on a unique Delphi consultation type, a model is proposed for supporting randomness in blockchain smart contracts. We understand, however, that although this was an incredibly intriguing and didactical philosophical exercise, implementing and supporting the recommendations provided in this research would require further research from a more technical perspective. We believe, however, that this work laid the groundwork for further interdisciplinary research in both classic and computer science, where professionals from both fields can contribute to developing innovative and groundbreaking oracle system designs.

**Funding.**

The authors received no external funding for this research.

**Author Contribution.**

G.C. Research idea, methodology, and writing (excluding sections 4 to 4.11)

M.O. Writing from sections 4 to 4.11

**Conflict of Interest.**

The authors declare no conflicts of interest.